\documentclass[a4paper]{jpconf}
\usepackage{graphicx}
\usepackage{wrapfig}

\begin{document}

\title{FRIGA, A New Approach To Identify Isotopes and  Hypernuclei In N-Body Transport Models.}
\author{\underline{A. Le F\`evre}$^{1}$, Y. Leifels$^{1}$, J. Aichelin$^{2}$, Ch. Hartnack$^{2}$, V. Kireyev$^{3}$ and E. Bratkovskaya$^{4}$\\
}
\address{
$^{1}$GSI Helmholtzzentrum f\"ur Schwerionenforschung, Darmstadt, Germany;\\ $^{2}$Subatech Nantes, France; 
$^{3}$JINR Dubna, Russia; $^{4}$FIAS and ITP Frankfurt, Germany.
}
\ead{A.LeFevre@gsi.de}

\begin{abstract}
We present a new algorithm to identify fragments in computer simulations of relativistic heavy ion collisions. It is based on the simulated annealing technique and can be applied to n-body transport models like the  Quantum Molecular Dynamics. This new approach is able to predict isotope yields as well as hyper-nucleus production. In order to illustrate its predicting power, we confront this new method to experimental data, and show the sensitivity on the parameters which govern the cluster formation.
\end{abstract}

\section*{Introduction}
In heavy ion reactions at energies between 20 A.MeV and several A.GeV, many clusters are formed. This cluster formation
presents a big challenge for transport models in which nucleons are the degrees of freedom which are propagated.
Identifying clusters in a transport code which transports nucleons is all but simple and therefore
in many approaches the fragment formation is simply omitted. This invalidates the prediction of single particle observables 
as well, because the cluster formation -- and therefore the modification of the single particle spectra due to the fragment formation -- depends on the phase space region and, 
as a consequence, cannot be approximated by a momentum independent scaling factor.  

The simplest way to identify clusters is by employing coalescence or a minimum spanning tree procedure. The first needs a
multitude of free parameters, whereas the second allows only for an identification at the end of the reaction which excludes any study on the physical origin \cite{gos97}. In addition, quantum effects,  like additional binding energies due to closed shells or pairing energies, are not supplied by the underlying transport theory which is semi-classical. 


\section*{The principles of the fragment recognition.}
If one wants to identify fragments early,  while the reaction is still going on,
one has to use the momentum as well as the coordinate space informations.
An idea how to do this has been launched  by Dorso et al. \cite{dor93}. It has been further developed into the
Simulated Annealing Clusterisation Algorithm (SACA) \cite{pur00} in the late 1990's and has been successfully applied to 
understand the measured fragment charge distribution and spectra as well as bimodality \cite{zbi07, lef09}. Starting from the positions and momenta of the nucleons at a given time during the reaction, nucleons are combined in all possible ways into fragments or single nucleons applying a simulated annealing technique. Neglecting the interaction among nucleons in different clusters, but taking into account the interaction 
among the nucleons in the same fragment, this algorithm identifies that combination of fragments and free nucleons
which has the \textit{highest binding energy}. If applied after the time when the energetic initial collisions are over, this most bound configuration has been proven to be close to the final distribution of fragments identified by the minimum spanning tree method at the end of the reaction\cite{pur00}.  The reason for this is the fact that fragments are not a random collection of nucleons  at the end, but an initial-final state correlation. SACA can be applied at any moment during the reaction and allows therefore for a detailed study of the fragment production mechanism.

In SACA, for accounting the interaction in-between nucleons, only the bulk Skyrme interaction supplemented by a Yukawa potential is used 
-- which is also the potential used for the propagation of the nucleons in the QMD transport model. 
To obtain more realistic fragment observables and to be able to predict observables for isotopes and
hyper-nuclei, we employ in our new approach a more realistic interaction and add the secondary decay because the fragments,
when identified, have a (moderate) excitation energy. This new approach is dubbed FRIGA ("Fragment Recognition In General Application").


\section*{The new features of FRIGA.}

\begin{wrapfigure}{r}{0.49\linewidth}
  \centering
   	\vspace{-10mm}
  \includegraphics[width=\linewidth]{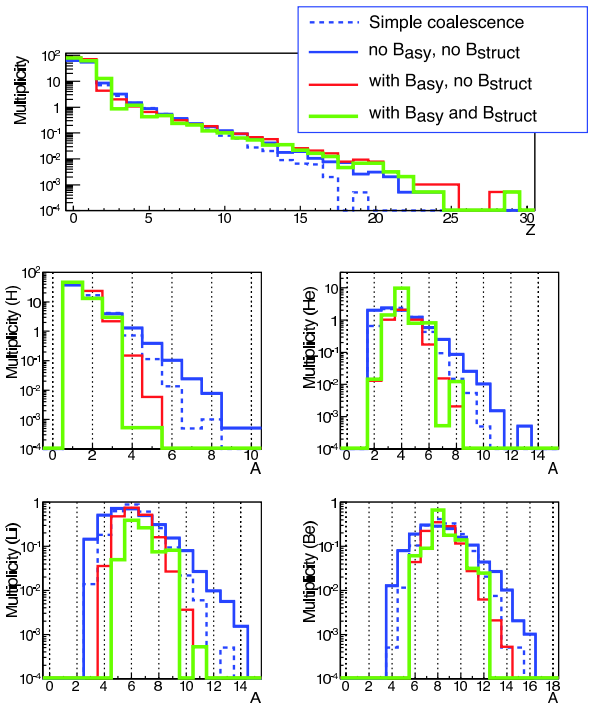}
   	\vspace{-7mm}
  \caption{\small{IQMD predictions for the central ($b<0.2 b_{max}$) collisions of
      $^{124}Xe+^{112}Sn$ at 100 A.MeV incident energy.
      Dashed line for
      the MST (coalescence) algorithm alone (performed at the late time 200
      fm/c), blue line for the initial SACA model, which has been
      extended into FRIGA with an asymmetry term (red) and additional nuclear structure
      contribution (green). The top panel shows the mean multiplicity
      distribution of fragments as a function of their charge. The four
      others depict the yields of  H, He, Be and Li isotopes.}}
  \label{fig1}
\end{wrapfigure}

In order to  predict the absolute multiplicity of the isotope yields, we have added new features to the SACA cluster identification.
They include the asymmetry energy, pairing and quantum effects. 

For the asymmetry energy, we adopt the parametrisation from  IQMD \cite{har98},  
a transport code which we use in the present article for the transport of
nucleons. For a proton the single particle energy thus reads:
\begin{displaymath}
  B_{asy}=E_{0}(\frac{<\rho_{B}>}{\rho_{0}})^{\gamma-1}  \frac{\rho_{n}-\rho_{p}}{\rho_{B}}
\end{displaymath}
where $E_{0}$=23.3 MeV, and
$\rho_{n}$, $\rho_{p}$, $\rho_{B}$, $\rho_{0}$ are the
neutron, proton, baryonic and saturation densities, respectively. In the present work, we take $\gamma$=1 (``stiff'' asymmetry potential).\\


Another significant part of the binding energy of light
isotopes are the shell structure and odd-even effects (pairing). In the
conditions of high pressure and temperature where FRIGA is used to
determine the pre-fragments, these structure effects are not well
known. E. Khan et al. \cite{kha07} showed that there are some indications 
that they affect the primary fragments. The authors
demonstrate that the pairing vanishes above a nuclear temperature $T_V\approx0.5\Delta_{pairing}$
 (pairing energy). At normal density the pairing energy tends to be
 negligible for heavy nuclei, with $\Delta_{pairing} = \frac{12}{\sqrt{A}}
 MeV$, whereas it is strong for light isotopes, like $^{4}He$ and $^{3}He$
 with 12 MeV and 6.9 MeV, respectively. In FRIGA, the primary fragments
 are usually produced slightly below the saturation density (typically around half of it) and 
 quite cold, with $T<1-2 MeV$, and hence below $T_V$.
Therefore, one cannot neglect the pairing energy. The same is true for shell effects
which produce experimentally a visible enhancement of the fragment yield for closed shell nuclei.

In order to determine the contribution of all structure effects to the binding energy of 
clusters, we make two hypotheses independent of the density and the average kinetic energy of the fragment environment.

First, the relative ratio of this nuclear structure contribution to the overall binding energy remains unchanged at the moderate temperatures and at  the density at which clusters are formed which is not far away from the saturation density. 

Introducing among the nucleons, initialised with the right root-mean-square radius , two body interactions, which corresponds in infinite matter to the Skyrme equation of state, the total fragment energy
\begin{eqnarray}
E_B(N,Z)= \langle H \rangle &=& \langle T \rangle + \langle V \rangle
\nonumber \\ 
&=& \sum_i \frac{p_i^2}{2m_i} +
\sum_{i} \sum_{j>i}
 \int f_i({\bf r, p},t) \,
V({\bf r, r\,', p, p\,'})  f_j({\bf r\,', p\,'},t)\, \rm d{\bf r}\, \rm d{\bf r\,'}
\rm d{\bf p}\, \rm d{\bf p\,'} \quad.
\end{eqnarray}
where $f_i$ is the single-particle Wigner
density
\begin{equation} \label{fdefinition}
 f_i ({\bf r, p},t) = \frac{1}{\pi^3 \hbar^3 }
 {\rm e}^{-\frac{2}{L} ({\bf r} - {\bf r_i} (t) )^2   }
 {\rm e}^{-\frac{L}{2\hbar^2} ({\bf p - p_i} (t) )^2 } \quad 
\label{binding}
\end{equation}
reproduces very well the nuclear binding energy given by the Weizs\"acker mass formula for ground state nuclei, $B_0$ \cite{aic91}. (Fig.12).

Our second hypothesis is that eq. \ref{binding} remains the right description of the binding energy if the nuclei are deformed or excited when the fragments are identified by the FRIGA algorithm. 

Taken both assumptions together, we can express  the nuclear structure contribution to the binding energy
 of a deformed cluster with Z protons and N neutrons  in the following way:
\begin{displaymath}
 B_{struct} = E_B(N,Z)  \frac{B_{exp}(Z,N)-B_{BW}(Z,N)}{B_{BW}(Z,N)}
 \end{displaymath} 
where $B_{exp}$ and $B_{BW}$ are the experimentally measured binding energy
(which contains the structure contribution  and the sum of the volume and
surface terms of the Bethe-Weizs\"acker formula). (Hyper-)Isotopes which are not stable at all in nature, are discarded in FRIGA by assigning to them a very
repulsive $B_{struct}$.  
The total binding energy of a cluster with N and Z, which is used in the annealing algorithm, will then be: 
\begin{displaymath}
 B = E_B(N,Z) + B_{asy} + B_{struct}. 
 \end{displaymath}
in contradistinction to SACA in which only the first term is used. 
 
Fig. \ref{fig1} shows the influence of the asymmetry energy and of the structure energy on the isotope yield
in the reaction $^{124}Xe+^{112}Sn$ at 100 A.MeV. We display here the results for central collisions ($b<0.2\ b_{max})$.
This figure illustrates as well how the various ingredients influence the  fragments yield  obtained in FRIGA, 
assuming an early clusterisation at t=60 fm/c. From that time on, the size of the pre-fragments does not change anymore. 
We see that the charge distributions are not strongly modified for the different options, whereas details of the isotopic 
yield are strongly influenced: the asymmetry energy tends to narrow the distributions towards the valley
of stability, whereas the structure effects contribute to restore the natural abundances, particularly strong for the $^{4}He$ clusters. 

\section*{Excitation energy and density of the primary fragments.}
The pre-fragments, called also ``primary'' fragments, created in
FRIGA, are often produced non relaxed in shape and density. When turning to their ground state, 
the shape surface energy is converted into excitation energy. Using QMD simulations,
for beam energies between 50 A.MeV - 1 A.GeV, FRIGA obtains for central heavy ion collisions a  mean excitation energy of the intermediate mass fragments  between 0 and 3 A.MeV, depending on the fragment size and
very similarly to the experimental measurements of \cite{hud03}. 
This excitation energy is sufficiently large that the secondary decay of the pre-fragments causes a significant contribution to the yield of small clusters.
For this reason, we optionally allow in FRIGA the excited cluster to undergo sequential secondary decays,
using the GEMINI algorithm \cite{char88}.  

Another interesting feature of the primary clusters in FRIGA is
their internal density. Although the medium is close to
$\rho_0$, at the stage of the collision when the primary cluster formation is stabilised,
just after the colliding system begins to separate, the fragments predicted by FRIGA
are produced quite dilute, typically around $\rho=\rho_{0}/2$ for intermediate mass fragments, and around $\rho=\rho_{0}/5$ for the light $Z<3$ isotopes. 
 This is explained by the fact that the dense clusters are disfavoured, 
because they would contain nucleons which are moving against each other. 
In this case the nucleons have a too high relative momenta to form a cluster.
Therefore, in the FRIGA approach, fragment formation tests only the low density behaviour of
the potentials, which are contributing to the binding energy. 

\section*{Another application of FRIGA: the hypernucleus formation.}

\begin{wrapfigure}{r}{0.49\linewidth}
  \centering
 \vspace{-6mm}
  \includegraphics[width=\linewidth]{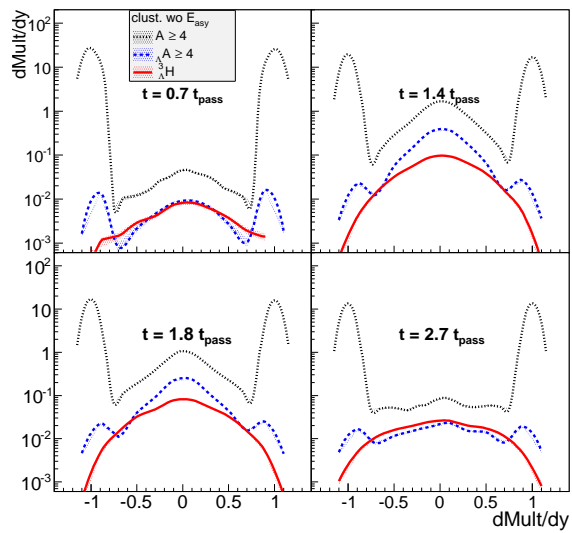}  
 \vspace{-7mm}
  \caption{\small{Predictions of FRIGA from PHSD \cite{cas09} simulations of Au+Au collisions at 11.45 A.GeV incident energy, b=6 fm.
  it shows the multiplicity per event as a function of the rapidity in the centre of the collision scaled to the projectile rapidity, for 
  various (hyper-)clusters ($A_\Lambda$ are hyper-clusters with $A\ge4$) and clusterisation times. $t_{pass}$=7.5 fm/c is the passing time of the projectile and target in central collisions.
  Like in the following figures, the shaded areas depict the statistical uncertainties.}}
  \label{fig2}
\end{wrapfigure}

An hyper-nucleus is a nucleus which contains at least one hyperon ($\Lambda(uds)$, ...) in addition to nucleons.
Extending FRIGA to the strange sector requires the knowledge of the $\Lambda$N potential.
In this first study, we consider the strange quark as inert and use 
$V_{\Lambda N} = \frac{2}{3} V_{nN}$ for protons as well as for neutrons.
Similarly, we consider the case of multiple strange nuclei as well, in which more than one hyperon is part of the fragment.
There, the coupling of 2  $\Lambda's$ contributes with the potential 
$V_{\Lambda\Lambda} = (\frac{2}{3})^{2} V_{nN}$. In the present approach
we neglect $B_{asy}$ for the hyperons, and take the contribution of the core nucleus (partner of the hyperons) 
as if it were decoupled from the hyperon.  Since the pairing and shell contributions in the binding energy are not yet well known for hyper-nuclei,  we neglect the $B_{struct}$  contribution. 

Using these modifications of the potentials, FRIGA produces hyper-nuclei with the same procedures as
non strange fragments. In the underlying QMD-like programs, which propagate
the hadrons, $\Lambda$'s are produced in different reactions:  
$\bar{K}+N\rightarrow\Lambda+\pi$, $\pi+n\rightarrow\Lambda+K^{+}$,
$\pi^{-}+p\rightarrow\Lambda+K_{0}$, $p+p\rightarrow\Lambda+X$.
Their abundance, position and momentum distributions are strongly
influenced by the reaction kinematics, the nuclear equation of state and the 
in-medium properties of the $K^+$ (kaon potential, etc.)
which are implemented in the transport model \cite{Hartnack:2011cn}.  

\begin{wrapfigure}{r}{0.49\linewidth}
  \centering
  \includegraphics[width=\linewidth]{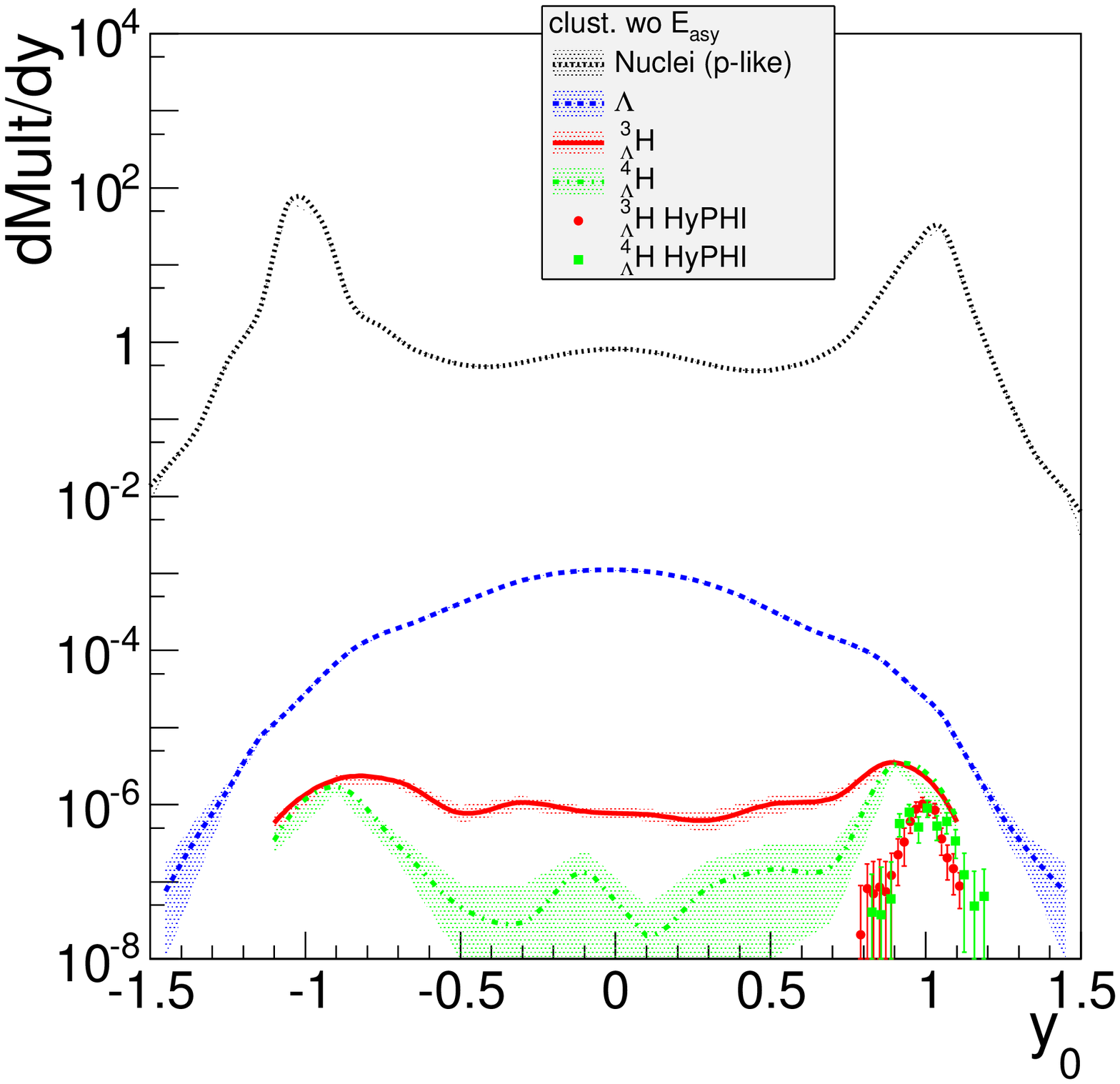}  
  \vspace{-9mm}
  \caption{\small{Predictions of FRIGA (clustering at $2 t_{pass}$, binding energy excluding $B_{asy}$) from IQMD simulations of $^{6}Li+^{12}C$ collisions at 2 A.GeV incident energy, $b>3 fm$
  compared to the HyPHI experimental data. The results of the model calculations are not filtered for the experimental acceptance. It shows the multiplicity per event per unit of rapidity, as a function of the rapidity in the centre of the collision scaled to the projectile rapidity, for 
  all clusters (in proton-like weighting), $\Lambda's$, ${}^{3}_{\Lambda}H$ and ${}^{4}_{\Lambda}H$. Markers are experimental data.}}
  \label{fig3}
  
  \includegraphics[width=\linewidth]{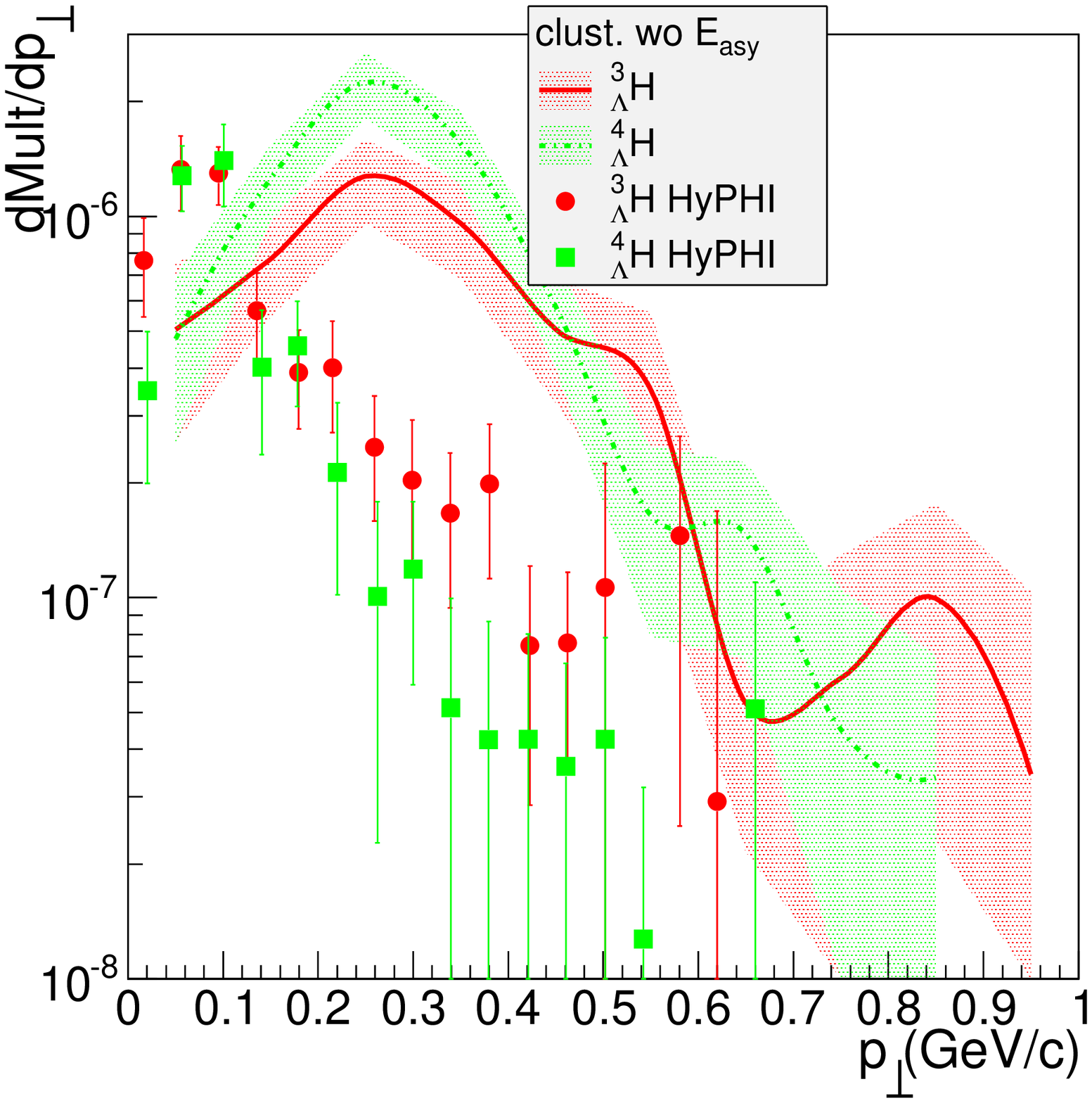}  
  \vspace{-10mm}
  \caption{\small{Same as Fig.~\ref{fig3}, showing the transverse momentum multiplicity distributions in the projectile spectator region ($y_{0}>0.9$). }}
  \label{fig4}
\end{wrapfigure}


Due to their composition, the yields of hyper-nuclei are produced when a cluster in coordinate and 
momentum space absorbs a hyperons. In heavy ions collisions at relativistic energies, 
the hyperon distributions are strongly peaked around the mid-rapidity region whereas the large 
fragments have rapidities close to the beam or target rapidity. The closer the rapidity of the hyperon approaches
-- by production or by subsequent collisions -- the target/beam rapidity, the larger is the probability that it can be absorbed by 
one of this larger clusters. Heavy hyper-nuclei are therefore observed not far away from beam/target rapidity.
At the same time hyperons can also form with other nucleons light clusters at mid-rapidity. There, the probability decreases
with the cluster size because it is increasingly difficult to form large cluster out of a gas of nucleons. Whereas the large
clusters in the beam/target rapidity regime can be identified quite early, the light clusters at mid-rapidty are formed later
and many of them decay due to the interactions with the surrounding nucleons which form a gas of a large temperature as compared to the cluster binding energy. 
This is illustrated in Fig.~\ref{fig2}.

As seen previously, in Fig.~\ref{fig1}, the ingredients of the cluster binding energy influence the light isotope yields in FRIGA. The
same is observed for hypernuclei. Adopting the factor $\frac{2}{3}$ in $V_{\Lambda N}$ has a strong effect, decreasing 
on the average the hypernucleus yields by around 20 percent. The asymmetry energy in the cluster can have a similar effect, depending
on the isotope (Z,N) asymmetry.

In order to illustrate the predicting power of the FRIGA algorithm, we confront it to experimental observations of light hypernuclei produced in the 
spectator region in collisions of the light system $^{6}Li + ^{12}C$ at 2 AGeV incident energy, measured by the HyPHI Collaboration 
at the SIS synchrotron of GSI Darmstadt. The data presented here are taken from \cite{rap15}.
Fig.~\ref{fig3} compares the IQMD-FRIGA predictions for the rapidity distributions of ${}^{3}_{\Lambda}H$ and ${}^{4}_{\Lambda}H$ with the experiment. 
The best agreement in the experimentally resolved rapidity region (close to the projectile spectator, $y/y_{beam}>0.7$) has been obtained while excluding the most central collisions (taking b$>$3 fm).
This procedure is a very basic approach to the simulate the effect of the complex experimental trigger.  
The chosen rapidity region has the highest hadronic yield and contains still  the tail of the $\Lambda$ distribution, as predicted by IQMD-FRIGA.
At these rapidities, the experiment has measured a yield ratio $Y({}^{3}_{\Lambda}H)/Y({}^{4}_{\Lambda}H) = 1.4 \pm 0.8$, with which IQMD-FRIGA agrees within the experimental uncertainty
with $1.3 \pm 0.2$. Including the asymmetry contribution $B_{asy}$ in the cluster binding energy in FRIGA, we obtain a yield ratio of $1.9 \pm 0.4$ which is still within this uncertainty. Therefore, at this level, the role of the asymmetry energy is difficult to judge.   
Fig.~\ref{fig4} shows that the transverse momentum distributions in the spectator region agree as well, in the slopes and the absolute yield. Just a slight discrepancy has to be noticed with a shift of 0.1 GeV/c in $p_\perp$. Here, we present the results without $B_{asy}$ contribution, but the conclusion is similar when including it. 

\section*{Conclusion}
We present here the first step towards an understanding of the production of isotopic yields and hypernuclei
in heavy ion reactions. 
Our clusterisation algorithm FRIGA, an improved version of the SACA approach,  which
 includes pairing and asymmetry energies as well as other structure effects  is able to describe more precisely the nuclear binding energy and
allows for realistic predictions of absolute (hyper-)isotope yields.
 We have seen that the asymmetry and pairing potentials can have a strong influence on
 both, the yields and momentum anisotropies for the
 (hyper-)isotopes. 
 According to this model, the nucleons which form fragments have 
initially a fairly low density. They contract and form
finally slightly excited fragments which may undergo secondary decays. 
Therefore, the fragment formation is sensitive to
the density dependence of the asymmetry energy and the pairing energy. However,
fragments test this dependence only for densities below the saturation density.

\end{document}